\documentstyle[aps,eqsecnum]{revtex}
\def\R2{I\!\!R}
\def\N2{I\!\!N}
\begin{document}
\title{\bf\huge COHERENT STATES \\ MEASUREMENT ENTROPY }
\author{{\Large Jaros{\l}aw Kwapie{\'n}$^1$, Wojciech
S{\l}omczy{\'n}ski$^2$ and Karol
\.Zyczkowski$^3$}  \\
\vspace{1.0cm}
         \begin{tabular}{lll}
$^1$ & Instytut Fizyki J{\c a}drowej im. H.
Niewodnicza{\'n}skiego, \\
  & ul. Radzikowskiego 152,
   PL-31 305, Krak\'ow, Poland\\
$^2$ & Instytut Matematyki, \\
   & Uniwersytet Jagiello\'nski,
        ul. Reymonta 4\\
   & PL-30 057, Krak\'ow, Poland\\
$^3$ & Instytut Fizyki im. M. Smoluchowskiego, \\
   & Uniwersytet Jagiello\'nski,
        ul. Reymonta 4 \\
   & PL-30 057, Krak\'ow, Poland\\
\end{tabular}
}
\maketitle
\vspace{0.9cm}
\begin{center}
{\small e-mail: $^1$kwapien@castor.if.uj.edu.pl \quad
$^2$slomczyn@im.uj.edu.pl \quad $^3$karol@castor.if.uj.edu.pl }
\end{center}
\vspace{1.2cm}
\begin{abstract}
Coherent states (CS) quantum entropy can be split into two components. The
dynamical entropy is linked with the dynamical properties of a quantum
system. The measurement entropy, which tends to zero in the semiclassical
limit, describes the unpredictability induced by the process of a quantum
approximate measurement. We study the CS--measurement entropy for spin
coherent states defined on the sphere discussing different methods dealing
with the time limit $n \to \infty$. In particular we propose an effective
technique of computing the entropy by iterated function systems. The
dependence of CS--measurement entropy on the character of the partition of
the phase space is analysed.
\end{abstract}

\newpage

\section{Introduction}

\label{s:introduction}
During the last decade a lot of attention has been
paid to the analysis of quantum analogues of chaotic classical maps defined
on a compact phase space. In particular quantum versions of the Baker map
\cite{BV89,S91}, the Arnold cat map \cite{HB80} (torus) and the periodically
kicked top \cite{HKS87,FM86,NOB86} (sphere) become standard models often
used in the study on {\sl quantum chaology} \cite{H91,G91,P93}. The
classical versions of these models can be called {\sl chaotic} (for the
kicked top under an appropriate choice of parameters), since the Kolmogorov
- Sinai (KS) dynamical entropy of the systems is positive. The definition of
KS--entropy cannot be adopted straightforwardly into quantum mechanics,
as it is based on the concept of classical trajectory. Several methods of
generalizing KS--entropy to quantum mechanics were proposed (see \cite
{W91,OP93,SZ94,R95,VM95} for the complete bibliography), but most of them
lead to zero entropy for all quantum systems represented on a
finite--dimensional Hilbert space. This property is common to the
definitions of quantum entropy due to Connes--Narnhoffer--Thirring \cite
{CNT87}, Gaspard \cite{Ga94}, Alicki--Fannes \cite{AF94}, and Roepstorff
\cite{R96}. Therefore these concepts of quantum entropy are not suitable to
describe dynamical properties of the above mentioned quantum maps which, due
to the compactness of the classical phase space, act on a
finite--dimensional Hilbert space.

{\sl Coherent states (CS) dynamical entropy}, introduced in \cite{SZ94},
appears to be more adequate for a quantitative
chara\-cte\-rization of chaos in
such quantum systems. This definition of quantum entropy takes into account
the process of {\sl sequential approximate measurement}. The notion of
approximate, or unsharp, or fuzzy quantum measurement has been analysed in
the last 25 years by Ali, Emchs and Prugove{\v {c}}ki, Busch and Lahti,
Davies and Lewis, Ozawa, Schroeck and many others in the framework of the
operational approach to quantum mechanics, or stochastic (phase space)
quantum mechanics (see monographs \cite{D76}, \cite{Pr86}, \cite{BLM91}, and
\cite{Sc95} for details and further references). In order to get some
information about the localization of a state in a phase space one may
perform double (or multiple, in a phase space of higher dimension) quantum
measurement of canonically conjugated observables. Due to the uncertainty
principle such a measurement cannot be sharp and has to be approximate. The
definition of CS--entropy is therefore based on the modified postulate of
wave function collapse. The original postulate of L{\"{u}}ders and von
Neumann, corresponding to a single exact measurement, assumes that after a
measurement the state undergoes a transition to an eigenstate of the
observable \cite{L51}. The modified postulate, used in the description of an
approximate multiple measurement, asserts that after the measurement the
state is transformed into an appropriate mixture of coherent states, i.e.,
the coherent states are {\sl a posteriori} states in the sense of Ozawa \cite
{O85}.

The CS-entropy with respect to a given partition can be divided into two
parts: the {\sl dynamical entropy} which describes the dynamical properties
of a quantum system and the {\sl measurement entropy} related to the
unpredictability induced by the process of sequential quantum measurement.
The proof of the fact that in the semiclassical limit the CS--measurement
entropy tends to zero was sketched in \cite{SZ94}. In the same paper it was
conjectured that {\sl the CS--dynamical entropy tends to the KS--entropy of
the corresponding classical system} if the unitary dynamics comes from an
appropriate quantization procedure, and some results in this direction were
obtained.

In this work we analyse in detail CS--measurement entropy for the SU(2)
coherent states, where the phase space is the two--dimensional sphere. Such
an example is of special physical interest, since it corresponds to an
unsharp measurement of the spin components \cite{Sc82,Bu86,BS89,G90}. In
\cite{SZ94} we proposed a general plan for studying the notion of
CS--entropy. The results obtained here constitute the first step towards the
realization of this scheme.

This paper is organized as follows. In sect.~II we recall the definitions of
CS--entropy, CS--measurement entropy and CS--dynamical entropy, and review
some of their basic properties. In sect.~III we summarize the standard facts
on the SU(2) vector coherent states. Various methods of computing
CS-measurement entropy for the partition of the sphere into two hemispheres
are presented, and the semiclassical limit is discussed in sect.~IV. The
case of an arbitrary partition of the sphere is analysed in sect.~V. In
sect.~VI we treat yet another method of calculating CS--entropy based on the
notion of iterated function systems. The R\'{e}nyi--type generalizations of
CS--entropy are introduced in sect.~VII. Finally, sect.~VIII contains some
concluding remarks.

\section{Coherent States Quantum Entropy}

\subsection{Kolmogorov-Sinai entropy}

Coherent states entropy can be regarded \cite{SZ94} as a generalization of the
classical Kolmogorov--Sinai entropy. Let us recall here the definition of
the KS--entropy for a classical map $S:\Omega \to \Omega $ generating a
discrete dynamical system. Let $\Omega $ be a compact phase space endowed
with a probability measure $\mu $ and divided into $k$ disjoint measurable
cells $E_1,\dots ,E_k$. The time evolution of classical trajectories during
$n$ periods is described via probabilities

\begin{equation}
P^{cl}_{i_0,...,i_{n-1}}=\mu(\{x\in\Omega: x\in E_{i_0}, S(x)\in E_{i_1},
\dots, S^{n-1}(x) \in E_{i_{n-1}} \}),
\label{pclas}
\end{equation}
of entering a given sequence of cells, where $i_l=1,\dots,k;~l=0,\dots,n-1$.
It is assumed that the initial points, determining uniquely each trajectory,
are distributed uniformly in the phase space with respect to the measure
$\mu $.

The partial entropy of $S$ is

\begin{equation}
H_n = - \! \! \sum_{i_0,\dots,i_{n-1}=1}^{k} P^{cl}_{i_0,\dots,i_{n-1}} \ln
P^{cl}_{i_0,\dots,i_{n-1}},
\label{partial}
\end{equation}
the {\sl KS--entropy with respect to the partition} ${\cal {C}}
=\{E_1,\dots,E_k\}$ is given by

\begin{equation}
H_{KS}(S,{\cal {C}}) := \lim_{n\to \infty} {\frac{1 }{n}} H_n,
\label{gran}
\end{equation}
and finally the {\sl KS--entropy} of $S$ is defined as \cite{ER85}

\begin{equation}
H_{KS}(S):=\sup_{{\cal {C}}}H_{KS}(S,{\cal {C}}).
\label{parks}
\end{equation}
In the above formula the supremum is taken over all possible finite
partitions of the phase space. A partition for which the supremum is
achieved is called {\sl generating}. Knowledge of a $k$--element generating
partition for a given map allows one to represent the time evolution of the
system in a $k$--letters symbolic dynamics and to find the upper bound for
the KS-entropy: $H_{KS}(S)\le \ln {k}$. For some classical systems, like the
Baker map, it is straightforward to find a generating partition and to
compute the KS-entropy. On the other hand, it is usually difficult to find a
generating partition for an arbitrary classical map. Recent years brought
some progress in this field: Christiansen and Politi found a good
approximation for a generating partition for the standard map \cite{CP95}
and obtained a fair estimate for the KS-entropy (see also \cite{ZS96}).

The convergence to the limit in (\ref{gran}) is usually slow, not faster
than ${1/n}$. It is therefore advantageous to consider the relative
entropies $G_n$ defined as

\begin{equation}
G_n:=H_n-H_{n-1},\ \hbox{for}\ n>1;\ G_1=H_1.
\label{gn}
\end{equation}
It is easy to show
 that the sequence $G_n$ also tends to $H_{KS}$ \cite{M87}.
This limit is achieved usually much faster than the limit in (\ref{gran}).
For example Misiurewicz and Ziemian \cite{MZ87,Z87} proved that for a
certain class of maps from the unit interval onto itself this convergence is
exponential (see also \cite{SG86}). It seems that such a behaviour is
typical for chaotic maps. We refer the reader to \cite
{Sz89,CGST93,EFR95,RFE96} for the review of recent results in this area.
Note that the convergence of ${\frac 1n}H_n={\frac 1n}\sum_{i=1}^nG_i$ is
slower, since the terms of larger $i$ have to balance a poor precision of
the approximation due to the initial terms \cite{CP85}.

\subsection{Approximate measurement and coherent states}

The probabilities $P^{cl}$ entering the definition of classical KS--entropy
(\ref{partial}) are meaningful under the assumption that during the
time evolution of the system one can trace an individual trajectory
and determine its localization in the phase space with infinite
precision. This supposition, consistent with the principles of
classical mechanics, is definitely not fulfilled in quantum
mechanics.

Information concerning the time evolution of a quantum system may be
obtained by the process of sequential measurement. The fundamental analysis
of a single quantum measurement of a discrete observable $\hat A$, expanded
in an orthonormal basis as $\hat A:=\sum_{m=1}^N a_m |m\rangle\langle m|,$
leads to the {\sl collapse postulate} of L{\"u}ders and von Neumann. The
canonical measurement of $\hat A$ yields with the probability
$p_a=\sum_{a_m=a} \langle m|\hat \rho|m \rangle$ the state reduction
\cite{L51}

\begin{equation}
\hat{\rho}~~-^{measurement}\hskip -1.9cm\!\!-\!\!\!-\!\!\!-\!\!\!-\!\!\!-\!
\!\!-\!\!\!-\!\!\!-\!\!\!-\!\!\!-\!\!\!-\!\!\!\to ~~\hat{\rho}^{\prime }:={
\frac{\sum_{a_m=a}|m\rangle \langle m|\hat{\rho}|m\rangle \langle m|}{
\sum_{a_m=a}\langle m|\hat{\rho}|m\rangle }},
\label{ludd}
\end{equation}
provided the outcome is $a$, where $\hat{\rho}$ is a density matrix
describing the state of the system before the measurement. If $|m\rangle $
is the only eigenstate of $\hat{A}$ corresponding to the eigenvalue $a_m$,
then this formula simplifies and the act of measurement transforms $\hat{\rho
}$ into a pure state $\hat{\rho}^{\prime }=|m\rangle \langle m|$.

The measurement of a single observable does not provide sufficient
information about the localization of the quantum state in the phase space.
Such information can be acquired only in a simultaneous double (or multiple)
approximate measurement of canonically conjugated observables. Let us
consider an $N$--dimensional complex Hilbert space ${\cal H}$ which
represents the {\sl kinematics} of the system and a compact set $\Omega$
equipped with a probability measure $\mu$ (we shall write $d x$ for
$d\mu(x)$) which forms a {\sl phase space}, or in the other words, a
space of experimental outcomes. A correspondence between both spaces
can be established by introducing a {\sl family of coherent states},
i.e., a continuous map {\hbox{$\Omega \ni x \longrightarrow |x \rangle
\in {\cal H}$} satisfying the resolution of the identity $\int_{\Omega}|x
\rangle\langle x|d x = I$ \cite{KS85}. In this work we use the coherent
states normalized as $\langle x|x \rangle=N$.

Following the ideas of Davies and Lewis \cite{DL70,D76} we assume, in a full
analogy to (\ref{ludd}), that a multiple approximate quantum measurement
yields the state reduction

\begin{equation}
\hat{\rho}~~ -^{measurement}\hskip -1.9cm \!\!-\!\!\!-\!\!\!
-\!\!\!-\!\!\!-\!\!\!-\!\!\!-\!\!\!-\!\!\!-\!\!\!-\!\!\! -\!\!\!\to ~~ \hat{
\rho}^{\prime}:= {\frac{{\frac{1 }{N}} \int_{E_i} |x \rangle \langle x |\hat{
\rho}|x \rangle\langle x| dx }{\int_{E_i} \langle x |\hat{\rho}|x \rangle dx
}},
\label{modludd}
\end{equation}
provided the outcome is in the cell $E_i$, which occurs with the probability
$P^{CS}_i = \int_{E_i} \langle x |\hat{\rho}|x \rangle dx$. Note that if one
increases the precision of the measurement of a single variable (and
simultaneously decreases the precision of the measurement of the canonically
coupled variable) this postulate reduces in the limit to the standard
collapse postulate of L{\"u}ders and von Neumann. Formally, one has to
replace the coherent states $|x \rangle$ used in (\ref{modludd}) by
so--called {\sl squeezed states}.

\subsection{CS--probabilities and CS--entropy}

Our approach to quantum entropy is based on the assumption that the
knowledge about the time evolution of a quantum state is obtained from a
sequence of multiple approximate quantum measurements. The evolution of the
system between every two subsequent measurements is governed by a unitary
matrix $U$.

A scheme of the first three periods of the time evolution of the dynamical
system is presented in Fig.1. Consider a quantum path encoded by the
following sequence of cells $\{W,Y,Z,\dots \}$. Let the initial state be
proportional to the identity operator, i.e., $\hat{\rho}_0= 1/N \cdot I $.
The coherent states collapse postulate (\ref{modludd}) allows us to
calculate the probability that a given sequence of symbols occurs
(\cite{SZ94}, see also \cite{O93}). Namely, we have

\begin{eqnarray}
P_{W}^{CS} = & \int_{W} dw \langle w|\hat \rho_0| w \rangle = \mu(W), \\
P_{W Y}^{CS} = & \int_{W} dw \int_{Y} dy {\frac{1 }{N}} |\langle y|U| w
\rangle |^2 , \\
P_{W Y Z}^{CS} = & \int_{W} dw \int_{Y} dy {\frac{1 }{N}} |\langle y|U|
w\rangle|^2 \int_{Z} dz {\frac{1 }{N}} |\langle z|U| y \rangle|^2 .
\label{cspr1}
\end{eqnarray}

For arbitrary $n$ measurements delivering results in the cells
$\{E_{i_0},E_{i_1},\dots, E_{i_{n-1}} \}$ the probabilities are
equal to

\begin{equation}
P_{i_0,\dots,i_{n-1}}^{CS} = \int_{E_{i_0}} dx_0 \cdots \int_{E_{i_{n-1}}}
dx_{n-1} \prod_{u=1}^{n-1} {K(x_{u-1},x_{u})},
\label{csprn}
\end{equation}
where $i_l=1,\dots,k;~l=0,\dots,n-1$ and the kernel $K$ is given by

\begin{equation}
K(x,y) = {\frac{1}{N}} |\langle y|U| x\rangle|^2.
\label{kernelK}
\end{equation}
We call them {\sl CS--probabilities}. Partition dependent, {\sl coherent
states (CS) entropy} $H^{CS}$ of a quantum map $U$ is defined likewise its
classical counterpart (\ref{gran})

\begin{equation}
H^{CS}(U,{\cal {C}}) := \lim_{n\to \infty} {\frac{1 }{n}} H_n(U,{\cal {C}}),
\label{csent1}
\end{equation}
where

\begin{equation}
H_n(U,{\cal {C}}) := - \sum_{i_0,\dots,i_{n-1}=1}^{k}
P^{CS}_{i_0,\dots,i_{n-1}} \ln P^{CS}_{i_0,\dots,i_{n-1}},
\label{csent2}
\end{equation}
and ${\cal {C}}=\{E_1,\dots,E_k\}$. In the semiclassical limit the
CS--entropy seems to tend to the KS--entropy, if the quantization procedure
is {\sl regular} \cite{SZ94}, i.e., if some assumptions linking the family
of quantum maps with the corresponding classical map are fulfilled.

Quantum CS-probabilities can be also used to define other quantities which
measure the randomness of the system (for a recent account of such concepts
see \cite{A84} and \cite{K95}) like R\'{e}nyi--type entropy of order $\beta$
which we shall analyse in sect.~VII. For some purposes, for instance, it
might be useful to define {\sl CS--inverse participation ratio } $\nu$

\begin{equation}
\nu(U,{\cal {C}}) := \sum_{i_0,\dots,i_{n-1}=1}^{k}
(P^{CS}_{i_0,\dots,i_{n-1}})^2.
\label{invpart}
\end{equation}
It is an analogue of a quantity often used in solid state physics to
describe localization of a wave function \cite{WW77}, since its inverse
gives the average number of occupied cells. It is linked to CS--R\'{e}nyi
entropy of order 2.

In the simplest case of the trivial dynamics the quantum map $U$ reduces to
the identity operator $I$. Even in this case the quantum entropy $H^{CS}$
does not vanish, since the coherent states are not orthogonal and do overlap
\cite{KS85}. The {\sl CS--measurement entropy} is given by \cite{SZ94,ZS95}

\begin{equation}
H_{mes}({\cal {C}}) := H^{CS}(U\equiv I, {\cal {C}})
\label{csmes}
\end{equation}
and depends on a family of coherent states in the phase space $\Omega$ and
on a finite partition ${\cal {C}}$.

The {\sl CS-dynamical entropy of a quantum map U with respect to a partition
}${\cal C}$ is defined as \cite{SZ94,ZS95}

\begin{equation}
H_{dyn}(U,{\cal {C}}) := H^{CS}(U,{\cal {C}}) - H_{mes}({\cal {C}}),
\label{dyn2}
\end{equation}
and partition independent {\sl CS--dynamical entropy} as \cite{S95}

\begin{equation}
H_{dyn}(U) := \sup_{{\cal {C}}} H_{dyn}(U,{\cal {C}}).
\label{dyn3}
\end{equation}

In the present paper we study CS--measurement entropy and its dependence on
a partition and the semiclassical parameter. This is a preliminary step to
calculating CS--dynamical entropy, which is defined as the difference of two
quantities. Moreover, the techniques we use in computing of CS--measurement
entropy can be also applied in the general case.

\subsection{Properties of CS--measurement entropy}

We now review some basic properties of CS--measurement entropy. Let us
assume that a finite partition ${\cal {C}}$ of the phase space $\Omega$ is
given. Let $H_n({\cal {C}})$ be defined by (\ref{csprn})--(\ref{csent2})
with $U = I$, and let $G_1({\cal {C}}) = H_1({\cal {C}})$; ~ $G_n({\cal {C}}
) = H_n({\cal {C}})-H_{n-1}({\cal {C}}), ~ \hbox{for} ~ n>1$. Then, applying
the general theory of entropy for random transformations \cite{ME81,K88}, we
obtain the following facts:

\vskip 0.2cm {\bf (1)} the sequences ${\frac{1 }{n}} H_n({\cal {C}})$ and
$G_n({\cal {C}})$ decrease with $n$ to $H_{mes}({\cal {C}})$;

\vskip 0.2cm {\bf (2)} if a partition ${\cal {C}}^\prime$ is finer than a
partition ${\cal {C}}$, then $H_{mes}({\cal {C}}^\prime) \ge H_{mes}({\cal {C
}})$.

\vskip 0.2cm Next, let us observe that the kernel $K$ which appears in (\ref
{csprn}) is bistochastic, i.e., $\int_\Omega K(x,{\bar{y}})dx=\int_\Omega
K({\bar{x}},y)dy=1$ for all ${\bar{x}},{\bar{y}}\in {\Omega }$. Let us
denote by $K_0$ the maximum of $K$. Then

\vskip 0.2cm {\bf (3)} the CS--measurement entropy fulfills the following
inequalities:

\begin{equation}
{\frac{1 }{n}} H_n({\cal {C}}) - {\frac{1 }{n}} \ln K_0 \le H_{mes}({\cal {C}
}) \le {\frac{1 }{n}} H_n({\cal {C}}),
\label{bound1}
\end{equation}
and, in consequence,

\begin{equation}
H_1({\cal {C}}) - \ln K_0 \le H_{mes}({\cal {C}}) \le H_1({\cal {C}})
\label{bound2}
\end{equation}
(for the proof see Appendix A).

\vskip 0.2cm Note that $H_1({\cal {C}})$ does not depend on the family of
coherent states but only on the measure $\mu$ and it is just the entropy of
the partition ${\cal {C}}$ with respect to the measure $\mu$. If $\Omega$ is
a Riemannian manifold and $\mu$ is the Riemannian measure on $\Omega$, then
one can deduce from (2) and (3) that $H_{mes}({\cal {C}})$ can be
arbitrarily large for a sufficiently fine partition ${\cal {C}}$.

It follows from (1) and (3) that

\vskip 0.2cm {\bf (4)} if $H_1({\cal {C}}) \neq H_{mes}({\cal {C}})$, then
the sequence ${\frac{1 }{n}} H_n({\cal {C}})$ converges to the entropy $
H_{mes}({\cal {C}})$ precisely as $n^{-1}$

(for the proof see Appendix B).

\vskip 0.2cm Another important property of CS--measurement entropy,

\vskip 0.2cm {\bf (5)} $H_{mes}({\cal {C}})$ tends to $0$ in the
semiclassical limit,

\vskip 0.2cm \noindent  was proved in \cite{SZ94} for SU(2) (spin) coherent
states. The decay seems to be rather slow. We shall try to evaluate its rate
in the sequel.

\subsection{Matrix form of CS--probabilities}

Let ${\cal C} = \{E_1,\dots,E_k\}$. We assume that the kernel $K$ entering
formula (\ref{csprn}) has the form

\begin{equation}
K(x,y) = \sum_{l,r=0}^M a_{lr} g_r(x) f_l(y), ~~~{\rm for}~~ x,y \in X,
\label{kernel1}
\end{equation}
where $a_{lr}\in \R2$, $f_l,g_r:\Omega\to \R2$ are continuous, for $
l,r=0,\dots,M$, and $f_0=g_0\equiv 1$ (in fact we can always present $K$ in
such a form if the family of coherent states comes from the canonical
group-theoretic construction (see \cite{P86,ZFG90,ZF95}) with the
finite-dimensional Hilbert space ${\cal {H}}$; then $M$ is an increasing
function of the dimension of the Hilbert space). Let us define matrices $
A=[a_{lr}]_{l,r=0}^M$ and $B(i)_{rl}=\int_{E_i} g_r(x)f_l(x) dx$ for $
l,r=0,\dots,M$, $i=1,\cdots,k$. Then the CS--probabilities are given by the
first element of the following matrix product:

\begin{equation}
P^{CS}_{i_0,\dots,i_{n-1}}=\bigl(B(i_{n-1})A B(i_{n-2})A\cdots A B(i_0)\bigr)
_{00}
\label{prob2}
\end{equation}
(the proof will appear in \cite{S96}). Now one can show that the family of
the CS--probabilities generates on the code space ${\cal S}^{\N2}$, where $
{\cal S} = \{1,\dots,k\}$, a shift--invariant measure, which is {\sl
algebraic} in the sense of Fannes {\sl et al.} (see \cite{FNS91}). Clearly,
the decomposition of the kernel $K$ is not unique. Moreover, the assumption $
f_0 = g_0 \equiv 1$ is too restrictive. In fact, to apply the matrix method,
it is enough to know that the constant function $1$ is a linear combination
of the functions $f_0,\dots,f_M$ \cite{S96}.

The above formula makes the calculation of the CS--entropy much easier.
Moreover, it is a starting point for the further investigation of entropy
utilizing the theory of iterated function systems. We present in sect.~VI
some results in this direction. For a fuller treatment we refer the reader
to \cite{S96}.

\section{Spin Coherent States}

The two--dimensional sphere $S^2$ can be considered as the phase space of
the periodically kicked top. This classical dynamical system is known to
exhibit chaos under a suitable choice of system parameters \cite{HKS87}. In
order to study a quantum analogue of this system it is convenient to
consider the operator of angular momentum $J$. Its three components $
\{J_x,J_y,J_z\}$ are related to the infinitesimal rotations along three
orthogonal axes $\{x,y,z\}$ in ${\R2}^3$ and fulfill the standard
commutation relations $[J_l,J_m]=i\varepsilon _{lmn}J_n,$ where $l,m,n=x,y,z$
and $\varepsilon _{lmn}$ represents the antisymmetric tensor (from now on we
put $\hbar =1$). The operators $J_{\pm }=J_x\pm iJ_y$ and $J_z$ are
generators of the compact Lie group $SU(2)$. The eigenvalues $j(j+1)$, $
j=0,1/2,1,3/2,\dots $, of the Casimir operator $J^2=J_x^2+J_y^2+J_z^2$
determine the dimension $N=2j+1$ of the Hilbert spaces ${\cal H}_N$ carrying
the representation of the group. Common eigenstates $|j,m\rangle ,\
m=-j,\dots ,j$, of the operators $J^2$ and $J_z$ form an orthonormal basis in
${\cal H}_N$.

The $SU(2)$ (spin) coherent states were introduced by Radcliffe \cite{R71}
and Arecchi {\sl et al.} \cite{A72}. For a thorough discussion we refer the
reader to \cite{KS85,P86,ZFG90,ZF95,H87,VS95}. The idea is the following.
Each point on the sphere labeled by the spherical coordinates $(\vartheta,
\varphi) $ corresponds to the {\sl $SU(2)$ coherent state} $| j, \vartheta,
\varphi \rangle$ generated by the unitary operator~ $R(\vartheta, \varphi) =
\exp\bigr[i\vartheta\bigl( \sin \varphi J_x- \cos \varphi J_y\bigl)\bigl]$
acting on the reference state $|j,j\rangle$. The natural projection $SU(2)
\to SO(3)$ relates with the operator $R(\vartheta, \varphi)$ the rotation by
the angle $\vartheta$ around the axis directed along the vector $(\sin
\varphi, - \cos \varphi, 0)$ normal to the $z$--axis and to the vector $
(\sin \vartheta \cos \varphi, \sin \vartheta \sin \varphi, \cos \vartheta)$
(see Fig.~2). The state $|j,j\rangle$, pointing towards the "north pole" of
the sphere, enjoys the minimal uncertainty, i.e., the expression
$\sum_{l=x,y,z} \Delta J_l^2$ takes in this state the minimal value $j$ (the
other possible choice of the reference state is $|j,-j\rangle$). More
precisely we put

\begin{equation}
|j, \vartheta, \varphi \rangle = \sqrt{2j + 1} \cdot \, R(\vartheta,
\varphi) |j,j\rangle.
\label{csdef}
\end{equation}
Using the stereographical projection $\gamma = \tan(\vartheta /2)\exp
(i\varphi)$ one can find a complex representation of the coherent state \ \ \
\ \ \ $|j,\gamma\rangle:=|j,\vartheta,\varphi\rangle$

\begin{equation}
|j,\gamma \rangle ={\frac{\sqrt{2j+1}}{(1+|\gamma |^2)^j}}\exp [\gamma
J_{-}]|j,j\rangle .
\label{gamrot}
\end{equation}
The prefactor $\sqrt{2j+1}$ introduced into the above formulae ensures the
coherent states identity resolution in the form

\begin{equation}
\int_{S^2}|j,\vartheta ,\varphi \rangle \langle j,\vartheta ,\varphi |\;d\mu
(\vartheta ,\varphi )=I,
\label{resol}
\end{equation}
where the Riemannian measure $\mu $ on $S^2$ is given by $d\mu =\sin
\vartheta d\vartheta d\varphi /4\pi $ and therefore does not depend on the
quantum number $j$. The norm of the coherent states changes with $j$ as $
|\langle j,\vartheta ,\varphi |j,\vartheta ,\varphi \rangle |=2j+1$, which
enables the respective {\sl Husimi--like distribution} $S^2\ni (\vartheta
^{\prime },\varphi ^{\prime })\longrightarrow |\langle j,\vartheta ,\varphi
|j,\vartheta ^{\prime },\varphi ^{\prime }\rangle |^2\in \R2$ of the
coherent state $|j,\vartheta ,\varphi \rangle $ to tend to the Dirac {
\hbox{$\delta
$--function}} as $j\to \infty $. Thus, after such a renormalization we can
treat the limit $j\to \infty $ as the {\sl semiclassical limit} \cite{ZF95}
or, in the other words, as the {\sl sharp-point limit} in the sense of
Schroeck \cite{Sc85}. If we had transformed spin coherent states in a
different way defining $||j,\sqrt{2j}\gamma \rangle :=(1+|\gamma
|^2)^{-j}\exp [\gamma J_{-}]|j,j\rangle $, we would have obtained the
canonical (harmonic oscillator) coherent states in the limit $j\to \infty $.
This kind of limit, however, is completely different from the semiclassical
limit we use in the present paper.

To simplify the notation in the following sections we shall omit the number $
j$ labelling coherent states $|\vartheta, \varphi\rangle$ or $|\gamma\rangle$.
Note that $S^2$ is isomorphic to the coset space $SU(2)/U(1)$, where $U(1)$
is the maximal stability subgroup of $SU(2)$ with respect to the state $
|j,j\rangle$, i.e., the subgroup of all elements of $SU(2)$ which leave $
|j,j\rangle$ invariant up to a phase factor. Hence the above construction
can be treated as a particular case of the general construction of
group-theoretic coherent states.

Expansion of a coherent state in the eigenbasis of $J^2$ and $J_z$ reads

\begin{equation}
|\gamma \rangle = \sqrt{2j + 1} \sum_{m=-j}^{m=j} \gamma^{j-m} (1+\gamma\bar
\gamma)^{-j} \Bigl[\Bigl( {2j \atop j-m }\Bigr) \Bigr]^{1/2}
|j,m\rangle.
\label{gamro2}
\end{equation}
In the spherical variables $(\vartheta,\varphi)$ this expansion takes the
form

\begin{equation}
|\vartheta ,\varphi \rangle =\sqrt{2j+1}\sum_{m=-j}^{m=j}\sin ^{j-m}({\frac
\vartheta 2})\cos ^{j+m}({\frac \vartheta 2})\exp \Bigl(i(j-m)\varphi \Bigr)
\Bigl[\Bigl({2j \atop j-m}\Bigr)\Bigr]^{1/2}|j,m\rangle .
\label{thetrot}
\end{equation}
The expectation values of the components of $J$ are

\begin{equation}
\langle j,\vartheta,\varphi|J|j,\vartheta,\varphi\rangle = j(2j+1)\bigl(
\sin\vartheta\cos\varphi,\sin\vartheta\sin\varphi, \cos\vartheta\bigr)
\label{expect}
\end{equation}
which establishes the link between the coherent state $|j,\vartheta,\varphi
\rangle$ and the vector $(\vartheta,\varphi)$ oriented along the direction
defined by a point on the sphere.

The infinite basis formed in the Hilbert space by the coherent states is
overcomplete. Two different $SU(2)$ coherent states overlap unless they
point towards two opposite poles on the sphere. Expanding two coherent
states in the $|j,m\rangle $ basis (\ref{thetrot}) we can calculate their
overlap as

\begin{equation}
{|\langle \vartheta^{\prime},\varphi^{\prime}|\vartheta,\varphi \rangle|^2}
= (2j+1)^2 \Bigl({\frac{{1+\cos\Xi} }{2}}\Bigr)^{2j},
\label{overla1}
\end{equation}
where $\Xi$ is the angle between two vectors on $S^2$ related to the
coherent states $|\vartheta, \varphi \rangle$ and $|\vartheta^{\prime},
\varphi^{\prime}\rangle$. Hence the transition kernel $K$ defined by (\ref
{kernelK}) takes (for $U=I$) the form

\begin{equation}
K((\vartheta,\varphi),(\vartheta^{\prime},\varphi^{\prime})) = {\frac{{
|\langle \vartheta^{\prime},\varphi^{\prime}|\vartheta,\varphi \rangle|^2} }{
{2j+1}}} = {{\frac{{2j+1} }{2^{2j}}} \Bigl[ 1+
\cos\vartheta\cos\vartheta^{\prime}+
\sin\vartheta\sin\vartheta^{\prime}\cos(\varphi-\varphi^{\prime})
 \Bigr]^{2j} }.
\label{overla2}
\end{equation}
The overlap decreases to $0$ with $j$ for $|\vartheta, \varphi \rangle \ne
|\vartheta^{\prime}, \varphi^{\prime}\rangle$ and sufficiently large $j$.

\section{Measurement entropy for two hemispheres}

We would like to compute the CS--measurement entropy for the case
corresponding to the physical process of simultaneous approximate
measurement of different spin components. Let us first consider the simplest
case, where the classical phase space $\Omega $ equal to the
two--dimensional sphere $S^2$ is divided into two hemispheres $
E_{+}=\{(\vartheta ,\varphi ):\varphi \in [0,2\pi ),\vartheta \in [0,\pi
/2]\}$ and $E_{-}=\{(\vartheta ,\varphi ):\varphi \in [0,2\pi ),\vartheta
\in (\pi /2,\pi ]\}$. The result of any measurement $i=\pm 1$ gives
information about the orientation of the spin.

\subsection{Transition probabilities}

The CS-transition probabilities $P^{CS}$ for the results
 $i_0,\dots ,i_{n-1}$
of $n$ consecutive measurements are obtained from (\ref{csprn}) and
 (\ref{kernelK})
 by setting the evolution operator $U$
  to be the identity and
taking the appropriate integration
domains. The explicit integral reads

\begin{equation}
P^{CS}_{i_0,\dots,i_{n-1}} = (4 \pi)^{-n} \int_{E_{i_0}}\!\!
\sin\vartheta_0
d\vartheta_0 d \varphi_0\! \cdots\! \int_{E_{i_{n-1}}} \!\!
\sin\vartheta_{n-1} d\vartheta_{n-1}
 d \varphi_{n-1}\! \prod_{u=1}^{n-1} {K
\bigl( (\vartheta_{u-1},\varphi_{u-1}),(\vartheta_u,\varphi_u)\bigr) }
\label{cspr2}
\end{equation}
where the kernel $K$ is given by (\ref{overla2}), and $i_u = \pm 1$ for $u =
0,\dots,n-1$. Straightforward integration allows one to obtain analytical
results for low values of $n$ and $j$, collected in Tab.~1.

In spite of the trivial dynamics ($U\equiv I$) the result of the first
measurement may differ from the second one, and consequently, all the
transition probabilities are nonzero. In the semiclassical limit $j\to
\infty $ the ''mixed'' transition probabilities (e.g. $
P_{+-}^{CS}=P_{-+}^{CS}$) vanish, while the survival probabilities (e.g. $
P_{++}^{CS}$, $P_{+++}^{CS}$) tend to $1/2$. The geometric symmetry of
reflection induces the invariance of the probabilities with respect to the
interchange of signs $(+\longleftrightarrow -)$. Moreover, due to the
time--reversal invariance, the CS--probability for any sequence of results
equals the CS--probability of the same sequence written in the reverse order
(e.g. $P_{++-}^{CS}=P_{-++}^{CS}$, $P_{+++-}^{CS}=P_{-+++}^{CS}$, $
P_{++-+}^{CS}=P_{+-++}^{CS}$). Observe, that for a given number of
measurements $n$, the probabilities for two sequences of results with the
same number of transitions are similar (e.g. for one transition: $
P_{+---}^{CS}\approx P_{++--}^{CS}$; for two transitions: $
P_{+--+}^{CS}\approx P_{++-+}^{CS}$). Direct integration of (\ref{cspr2})
does not allow one to obtain the CS--probabilities for larger values of $j$
or $n$, which is necessary to estimate the CS--measurement entropy. For this
purpose it is convenient to formulate integrals in matrix form.

\subsection{Matrix formulation of integrals}

Computation of the CS-probabilities can be significantly simplified by
applying the general method described in sect.~II.E. This can be seen,
especially, for the division of the sphere into several latitudinal
components $E_1,\dots ,E_k$, where $E_i=\{(\vartheta ,\varphi ):\varphi \in
[0,2\pi ),\vartheta \in W_i\}$ for $i=1,\dots ,k$, and $\{W_1,\dots ,W_k\}$
is a partition of the interval $[0,\pi ]$. Performing the substitutions $
t_i=\cos \vartheta _i$ and integrating over $\varphi _0,\dots ,\varphi _n$
we can simplify formula (\ref{cspr2}) writing

\begin{equation}
P^{CS}_{i_0,\dots,i_{n-1}}= \int_{\tilde{W}_{i_0}}\! {\frac{1}{{2}}} dt_0 \
\cdots \int_{\tilde{W}_{i_{n-1}}}\! {\frac{1}{{2}}} dt_{n-1} \!
\prod_{u=1}^{n-1} {\tilde{K}\bigl(t_{u-1},t_u\bigr)},
\label{prob4}
\end{equation}
where $\tilde{W}_i=\{ \cos t: t\in W_i\}$ for $i=1,\dots,k$, and the reduced
kernel $\tilde{K}$ is given by

\begin{equation}
\tilde{K} \bigl( t,s \bigr) = {\frac{{2j+1} }{4^{2j}}} \sum_{q=0}^{2j} {
\Bigl({2j  \atop q} \Bigr)}^2 ((1+t)(1+s))^{q}
\bigl((1-t)(1-s)\bigr)^{2j-q} = \sum_{l,r=0}^{2j} a_{lr}t^l s^r,
\label{ker2}
\end{equation}
for $t,s \in [-1,1]$. Thus the kernel $\tilde{K}$ is represented in
the form (\ref{kernel1}) with $\tilde{\Omega} = [-1,1]$, $d\tilde{\mu}(t) =
{\frac{1}{2}} dt$, $f_l(t)=t^l$, $g_r(s)=s^r$ for $t,s \in \tilde{\Omega}$,
and $M = 2j$. Note that $\{\tilde{W}_1,\dots,\tilde{W}_k\}$ forms a
partition of $\tilde{\Omega}$. Hence we can apply formula (\ref{prob2}) for
the CS--probabilities writing them in the matrix form

\begin{equation}
P^{CS}_{i_0,\dots,i_{n-1}} = \langle (1,0,\dots,0)~| \bigl(B(i_{n-1})A
B(i_{n-2})A\cdots A B(i_0)\bigr)| (1,0,\dots,0) \rangle,
\label{prob3}
\end{equation}
with $A=[a_{lr}]_{l,r=0}^{2j}$ (given by (\ref{ker2})) and $B(i)_{rl} = {
\frac{1}{2}}{\int_{\tilde{W}_i}t^{l+r}dt}$ for $i=1,\dots,k$; $l,r =
0,\dots,2j$.

If we divide the sphere into two hemispheres, then $
B(i)_{rl}=i^{l+r}/2(l+r+1)$ for $i=\pm 1$; $l,r = 0,\dots,2j$. In this case
(\ref{prob3}) takes a particularly simple form for $j=1/2$

\begin{equation}
P^{CS}_{i_0,\dots,i_{n-1}} = {\frac{1}{2^n}} \big\langle (1,0)~\big| \left(
\matrix{1 & i_0/2 \cr i_0/2 &1/3\cr}\right) \left(\matrix{1 & i_1/2 \cr
i_1/2 &1/3\cr}\right) \cdots \left(\matrix{1 & i_{n-1}/2 \cr i_{n-1}/2
&1/3\cr}\right) \big| (1,0) \big\rangle,
\label{mat2}
\end{equation}
where the results of the measurements $i_u$ are equal to $-1$ or $+1$ for $
u=0,\dots,n-1$.

\subsection{Limit $n\to \infty$}

In the remainder of this section we assume that ${\cal C}$ is the partition
of the sphere into two hemispheres, i.e.,
 \ \ \hbox{${\cal {C}} = \{E_+,E_-\}$}.
Moreover, we set $H_{mes} := H_{mes}({\cal {C}})$, $H_n := H_n({\cal {C}})$
and $G_n := G_n({\cal {C}})$. In Tab.~2 we present partial and relative
entropies calculated for two different values of $j$ with the aid of the
formulae (\ref{csent1}), (\ref{csent2}) and (\ref{prob3}).

We assert in sect.~II.D(4) that $H_n$ converges to $H_{mes}$ exactly as $
\frac 1n$. One can deduce from Tab.~2 that the convergence of $G_n$ to the
same limit is much faster. In fact, it seems to be exponential. In sect.~VI
we give some arguments supporting this statement. Thus, to calculate the
limiting value we use the extrapolations ${H_n}\sim H_{mes}+\alpha /n$ and
${G_n}\sim H_{mes}+\gamma c^n$. The outcomes are contained in
Tab.~2. Let us observe that the rate of convergence decreases with
$j$ and hence the method of computing the CS-measurement entropy
based on formula (\ref{prob3}) does not lead to satisfactory
results in the semiclassical limit, i.e., for large quantum number
$j$.

\subsection{Semiclassical regime $j>>1$}

The matrix formula for the CS-probabilities is useful in numerical
calculations, but as it was mentioned above, do not allow us to compute the
entropy for very large values of $j$. For two measurements, however, one can
obtain some exact results. Applying (\ref{prob4}) and (\ref{ker2}) we get an
analytical formula for the CS--probability valid for any $j$

\begin{equation}
P^{CS}_{+-}= \Bigl({4j+1 \atop 2j}\Bigr) 2^{-4j-2}
\label{nn22}
\end{equation}
(for the proof see Appendix C).

Due to symmetry $P^{CS}_{-+}=P^{CS}_{+-}$ and $
P^{CS}_{++}=P^{CS}_{--}=1/2-P^{CS}_{+-}$. It is convenient to introduce a
$j$--dependent coefficient $\tau_j=P^{CS}_{+-}/P^{CS}_{++}$, which tends to 0
in the semiclassical limit $j \to \infty$. Using formula (\ref{nn22}) we
obtain

\begin{equation}
\tau_j = {\frac{\Bigl( {4j+1 \atop \ 2j} \Bigr)}{{2^{4j+1}- \Bigl(
{4j+1 \atop \ 2j} \Bigr)}}}.
\label{tau}
\end{equation}

In order to get an upper bound for the CS-measurement entropy we may compute
the relative entropy $G_2=H_2-H_1$ (see sect.~II.D). The partial entropy
after one measurement $H_1$ equals $\ln{2}$, independently of $j$. Summing
over four possible paths $++,+-,-+,--$ one can compute the partial entropy $
H_2$ obtaining finally a formula

\begin{equation}
G_2 = \ln(\tau_j+1) - {\frac{\tau_j }{\tau_j+1}} \ln(\tau_j),
\label{g2j}
\end{equation}
which is symmetric with respect to an involution $\tau_j \to 1/\tau_j$.

Inserting the expression (\ref{tau}) into the above formula we get an
explicit approximation for $H_{mes}$. It is represented by a solid line in
Fig.~3, while circles denote the results obtained numerically for small $j$
with the help of the matrix method presented above. In the semiclassical
range $j>>1$ it is legitimate to apply the Stirling approximation of the
factorial in (\ref{tau}), which gives

\begin{equation}
H_{mes}< G_2 \sim {\ln {\sqrt {2 \pi j}}} - \bigl(1-{\frac{1 }{\sqrt{2 \pi j}
}}\bigr) \ln(\sqrt{2 \pi j}-1)
\label{ggg}
\end{equation}
This formula, providing a fair approximation (and an upper bound) for the
CS--measurement entropy, is characterized by the asymptotic behaviour

\begin{equation}
G_2 \sim {\frac{\ln j }{2\sqrt{2 \pi j}}}.
\label{gg2}
\end{equation}

It is worth to note that formula (\ref{g2j}) can be obtained from the
Markovian approximation of the CS--probabilities. Let us assume for a moment
that the probabilities $P^{CS}_{i_0,\dots,i_{n-1}}$ were generated by a
Markov shift. It follows from the symmetry of the problem that its initial
vector would be $(1/2,1/2)$ and its transition matrix $Q$ would have the form

\begin{equation}
Q=\left(
\begin{array}{cc}
a & 1-a \\
1-a & a
\end{array}
\right), ~~~~~~~ {\rm where} ~~~ a=2P^{CS}_{++}={\frac{1 }{\tau_j+1}}.
\label{mark}
\end{equation}

In fact, our probabilities $P^{CS}_{i_0,\dots,i_{n-1}}$ are not generated by
a Markov shift, nevertheless, one can consider the Markovian approximation
as above. Then the approximate probabilities $P^{Mar}$ depending only on the
number $L$ of ``transitions'' from one hemisphere to the other
($L={\frac{1}{2}} (n-1-\sum_{q=1}^{n-1} i_q i_{q-1})$) are equal to

\begin{equation}
P_{i_0,\dots ,i_{n-1}}^{Mar}={\frac 12}Q_{i_0,i_1}\cdot \dots \cdot
Q_{i_{n-2},i_{n-1}}={\frac 12}(1-a)^La^{n-1-L}={\frac{\tau _j^L}{2(\tau
_j+1)^{n-1}}}.
\label{mar2}
\end{equation}
In this approximation the probabilities form a geometric series with the
same ratio $\tau _j$ for any number of measure\-ments~$n$. The precision of
such a Markovian approximation can be directly checked in Tab.~1 containing
the exact probabilities~$P^{CS}$. For example for $n=4$ and $j=1/2$ the
numerators $289,167(161),103(97)$ and $65$ form an approximately geometric
series, however, probabilities with the same number of the transitions are
not equal (e.g. $P_{+-++}^{CS}\ne P_{+--+}^{CS}$).

Summing over all $2^n$ possible sequences we obtain the following
approximate formula for the partial entropy:

\begin{equation}
H_n^{Mar} \!=-\sum_{L=0}^{n-1} \Bigl( {n-1  \atop L}
\Bigr) {\frac{\tau_j^L }{(\tau_j+1)^{n-1}}} \ln\bigr[{\frac{\tau_j^L }{2
(\tau_j+1)^{n-1}}}\bigl]=\ln2 + (n-1)\Bigl[ \ln(\tau_j+1) - {\frac{\tau_j }{
\tau_j+1}} \ln(\tau_j)\Bigr].
\label{mar3}
\end{equation}
Now dividing both sides by $n$ and performing the limit $n\to\infty$ we
arrive at the relative entropy $G_2$ given by (\ref{g2j}).

Let us recall that in the semiclassical limit $(j\to\infty)$ the relative
measurement entropy $G_2$ tends to zero as ${\rm ln} j/\sqrt{j}$. This
defines the scale in which the quantum effects reveal. Unfortunately the
precision of this approximation is not sufficient to conclude, whether the
logarithmic prefactor describes correctly the decay of the measurement
entropy $H_{mes}$ in the semiclassical limit, or whether its existence is an
artifact introduced by the approximation.

\section{CS--Measurement Entropy for Various Partitions}

The CS--measurement entropy depends on the number of cells in a partition
and on their shape. In this section we consider several partitions of a
different type: into two cells: both connected, one connected and one
disconnected, both disconnected, and into many cells. Figs.~4,~7,~and~10
contain schemes for these partitions. In all the cases we compute the
entropy using the matrix formulation introduced in sects.~II.E and IV.B. As
in the preceding section we assume that ${\cal C}$ denotes the respective
partition of the sphere, putting $H_{mes} := H_{mes}({\cal {C}})$, $H_n
:= H_n({\cal {C}})$, and $G_n := G_n({\cal {C}})$.

\subsection{Two connected cells}

Let us split the sphere into two segments along a parallel $\Theta_c$. The
northern segment $E_+$ contains points with $\vartheta\in[0,\Theta_c]$,
while the southern $E_-$ those with $\vartheta\in(\Theta_c,\pi]$. This
partition is shown schematically in Fig.~4.

Fig.~5 represents the dependence of the partial entropy $H_n/n$ on the
variable $\cos \Theta _c$ for $n=8$ measurements and several values of $j$.
For each value of $n$ and $j$ the partial entropy achieves its maximum at $
\cos \Theta _c=0$, for the partition into two hemispheres. The solid
horizontal line drawn at $\ln {2}$ represents the maximal entropy admissible
for the partition containing two cells. For increasing values of $j$ the
partial entropy decreases and tends to zero for $j\to \infty $.

For any of these partitions the partial entropy $H_n/n$ approaches the
limiting value $H_{mes}$ approximately as $1/n$ (see sect.~II.D(4)). As in
the previously discussed case of two hemispheres, we estimate the limiting
value $H_{mes}$ by computing the relative entropy $G_n$. Fig.~6 shows a
comparison of the partial entropies $H_n/n$ with $H_{mes}$ extrapolated in
this way for $j=1/2$, $n=2$ (Fig.~6a) and $j=5$, $n=2$ or $n=8$ (Fig.~6b).
The difference increases with the spin length $j$.

\subsection{Two disconnected cells}

Let us now analyse another two classes of partitions of the sphere into two
cells. In the first case (Fig.~7a) we divide the sphere into three parts
along parallels $\pi -\Theta _d$ and $\Theta _d$, and then join the lower
and upper parts, thus obtaining two cells: a connected spherical zone and
the disconnected union of two spherical segments. The CS--measurement
entropy (Fig.~8) changes in this case from $0$ (for $\Theta _d=0$) to $\ln {2
}$ (for $\Theta _d={\pi }/3$), which is the largest possible value for the
CS--measurement entropy with respect to a two-element partition.

In the second case (Fig.~7b) we start from the splitting of the sphere into
the lower and upper hemispheres. Next, we cut symmetrically two ``pieces of
cake'' out of both hemispheres, and then join the four parts across. We get
in this way two disconnected cells marked in black and white in Fig.~7b. The
CS--measurement entropy (Fig.~9) changes from $\ln {2}$ (for $\Phi _d=\pi $)
to $0.6613\dots $ (for $\Phi _d=0$). The latter case relates to the
partition of the sphere into two hemispheres studied in sect.~IV.

It is worth pointing out that in both cases the entropy attains the maximal
admissible value $\ln{2}$ for some value of the parameter. We do not know
whether one can generalize this observation and find a partition of an
arbitrary number of cells $k$ giving the maximal allowed value of the
entropy equal $\ln{k}$.

\subsection{Many cells}

Let us consider $k$ disjoint zones created on the sphere by $k-1$ parallels.
As in the case of two cells, represented in Fig.~4, the CS--measurement
entropy seems to achieve its maximum, if the cells have the same volume
$=1/k$ (see Fig.~10). We computed the CS--measurement entropy
$H_{mes}({\cal {C}}_k)$ for the partitions ${\cal {C}}_k$ of the sphere into
$k=2,\dots ,1000$ zones of the same volume. Note that for large $k$ even the
second relative entropy $G_2({\cal {C}}_k)$ provides a reliable estimate for
$H_{mes}({\cal {C}}_k)$. In Fig.~11 we present the CS-measurement entropy
displayed for $j=1/2$ as a function of the number of cells $k$ (circles). The
solid line represents the function $\ln {k}$, which gives the upper bound for
the entropy with respect to a partition consisting of $k$ cells.

Since for any partition ${\cal {C}}$ and a quantum map $U$ the CS--dynamical
entropy is defined by (\ref{dyn2}) as the difference of $H(U,{\cal {C}})$ and
$H_{mes}({\cal {C}})$, it is convenient to consider the quantity
$H_{max}({\cal {C}}_k):=\ln {k}-H_{mes}({\cal {C}}_k)$, limiting the partition
dependent dynamical entropy $H_{dyn}(U,{\cal {C}}_k)$ from the above. From
(\ref{bound2}) and (\ref{overla2}) we know that $H_{max}({\cal {C}}_k)\le
\ln ({2j+1)}$ as $H_1({\cal {C}}_k)=\ln {k}$. Although using this method one
can establish the finiteness of the partition independent dynamical entropy
$H_{dyn}(U)$ given by (\ref{dyn3}), this upper bound seems to be rather crude.
In fact $H_{max}({\cal {C}}_k)$ decreases with $j$. It is interesting to
observe that this quantity converges for $k\to \infty $. The limiting value
depends on $j$ and is close to $0.06$ for $j=1/2$. The inset in Fig.~11 shows
the dependence of $H_{max}({\cal {C}}_k)$ on $k^{-2}$ for $k=30,\dots ,1000$;
$j=1/2$. The data displayed in this way are well fitted by a straight line,
which allows us to postulate an approximate relation $H_{mes}(C_k)\approx \ln
{k}-0.05999745+0.1637/k^2$, found for $j=1/2$.

\section{CS-entropy and iterated function systems}

In this section we establish a relationship between CS--entropy and iterated
function systems (IFSs). Firstly, we show how to obtain an IFS from a
bistochastic kernel and a partition of the phase space. Then, we use this
system to get an integral formula for CS--entropy and propose a new method
of computing CS--entropy based on the ergodic theorem for IFSs. For more
information on IFSs see \cite{IG90}, \cite{B88}, and \cite{B89}.

\subsection{Iterated function systems and an integral formula for
CS--entropy}

We follow the notation of sects.~II.D and II.E. With each cell $E_i$ ($
i=1,\dots,k$) of the partition we associate an $(M+1) \times
(M+1)$~--~matrix\ $D(i) = B(i)A$. We consider functions $p_i : {\R2}^M
\to {\R2}^+$ and partial maps $F_i : {\R2}^M \to {\R2}^M$ given by

\begin{equation}
p_i(\lambda) = (1,0,\dots,0)(D(i)(1,\lambda))
\label{IFS1}
\end{equation}
and

\begin{equation}
F_i(\lambda) = (D(i)(1,\lambda))/p_i(\lambda)
\label{IFS2}
\end{equation}
for $\lambda \in {\R2}^M$, $i=,1\dots,k$.

Let us suppose that the functions $g_0 \equiv 1,g_1,\dots,g_M$ are lineary
independent. Then one can show that

\vspace{1.5mm} {\bf (0)} $p_i(\lambda) \ge 0$ for $i=1,\dots,k$ and
$\sum_{i=1}^{k} p_i = 1$, i.e., the functions $\{p_i\}_{i=1,\dots,k}$ can be
treated as place-dependent probabilities.

\vspace{1.5mm} Moreover we shall assume that there exists a set $X \subset {
\R2}^M$ such that

\vspace{1.5mm} {\bf (a)} X is a compact set with $\lambda_0 := (\int_X
g_1, \dots, \int_X g_M) \in X$,

\vspace{1mm} and for every $i=1,\dots,k$:

\vspace{1.5mm} {\bf (b)} $F_i(X) \subset X$,

\vspace{1.5mm} {\bf (c)} $p_i|_X > 0$,

\vspace{1.5mm} {\bf (d)} $F_i|_X$ is a Lipschitz function with the Lipschitz
constant $c_i < 1$.

\vspace{1.5mm} Then the following assertions hold:

\vspace{1.5mm} {\bf (1)} ${\cal F} = (F_i,p_i)_{i=1}^{k}$ is an {\sl
iterated function system} on $X$.

\vspace{1.5mm} {\bf (2)} The IFS ${\cal {F}}$ generates the following
operator $V$ acting on $M(X)$ (the space of all probability measures on X):

\begin{equation}
(V\nu)(B) = \sum_{i=1}^{k} \int_{F_i^{-1}(B)} p_i(\lambda) d\nu(\lambda)
\label{IFS3}
\end{equation}
for $\nu \in M(X)$ and $B \in B(X)$, where $B(X)$ denotes the family of all
Borel sets on $X$. This operator describes the {\sl evolution of probability
measures} under the action of ${\cal {F}}$. We shall denote by
$(Z_n^{\nu})_{n \in {\N2}}$ the associated
Markov stochastic process having the initial distribution $\nu$.

\vspace{1.5mm} {\bf (3)} There is a unique {\sl invariant probability
measure} $\mu$ for the IFS defined above fulfilling the equation $V\mu =
\mu$. This measure is attractive, i.e., $V^n{\nu}$ converges weakly to $\mu$
for every $\nu \in M(X)$ as $n \to \infty$.

\vspace{1.5mm} {\bf (4)} The relative entropies $G_n$ are given by

\begin{equation}
G_n = \int_{X} h_k(p_1(\lambda),\dots,p_k(\lambda))
d(V^{n}\delta_{\lambda_0})(\lambda) ~~\hbox{for} ~~n \in \N2,
\label{IFS4}
\end{equation}
where $h_k$ is the Shannon--Boltzmann entropy function given by $
h_k(p_1,\dots,p_k) = -\sum_{i=1}^{k} p_i \ln{p_i}$ ~for any $p_i \ge 0$
such that ~$\sum_{i=1}^{k} p_i = 1$.

\vspace{1.5mm}
{\bf (5)} The CS--entropy $H_{mes}$ is given by an {\sl integral formula}

\begin{equation}
H_{mes} = \int_{X} h_k(p_1(\lambda),\dots,p_k(\lambda)) d\mu(\lambda).
\label{IFS5}
\end{equation}

Let us sketch briefly the proof of the above statements. Assertion (1)
follows from (0) and assumption (b). The Markov processes generated by IFSs
were analysed in \cite{IG90} and \cite{BDEG88}. Assertion (3) can be deduced
from assumptions (c) and (d), and [65, Th.2.1]. Formulae (\ref{IFS4})
and (\ref{IFS5}) were proved by Fannes {\sl et al.} in \cite{FNS91} for
algebraic measures, i.e., under the assumption that the formula for
probabilities analogous to (\ref{prob2}) holds. They followed an earlier
result of Blackwell \cite{B57} on the entropy of functions of a finite--state
Markov chain. In these both papers, however, the authors did not refer to the
theory of IFSs and assumed that the matrices $D(i)$ are positive. In spite of
this, their proof can also be applied in our case. For more details we
refer the reader to \cite{S96}.

\subsection{Ergodic theorem and random algorithm for computing CS--entropy}

The key point in our reasoning is to find a set $X$ fulfilling conditions
(a)--(d) above. In all the cases we analysed this task was not too difficult
to accomplish. We shall give some examples below. Utilising the results
presented in \cite{IG90} and \cite{LY94} we can go even further and prove
(under some additional assumptions) that $G_n$ tends to $H_{mes}$
exponentially. Moreover, applying the {\sl Kaijser--Elton ergodic theorem
for IFSs} (see \cite{K81} and \cite{E87}) we obtain the following formula:

\begin{equation}
H_{mes} = \lim_{n\to \infty} {\frac{1 }{n}}\sum_{l=0}^{n-1} h(Z_l^{\nu}) ~~~~
{\mu}-{\hbox{almost everywhere}},
\label{IFS6}
\end{equation}
where $h = h_k(p_1,\dots,p_k)$ and $\nu$ is an arbitrary initial
distribution.

This formula gives another numerical method of computing CS--entropy. To
obtain the value $H_{mes}$ it suffices to calculate Ceasaro means of the
function $h$ along a trajectory of the stochastic process $(Z_l^{\nu})_{l
\in \N2}$. This is a particular case of the general method which appeared
under the name of Random Iterated Algorithm in \cite{B89}. The convergence
in (\ref{IFS6}) seems to be rather slow,
 namely as $n^{1/2}$. Note, however,
that here the time computational complexity grows with $k$ (the number of
elements of the partition) lineary, whereas in the ``matrix method'' we
considered in sects.~II.E~and~IV.B it grows polynomially (as $k^n$). Hence
the method based on formula (\ref{IFS6}) may be specially useful for
large values of $k$.

\subsection{Example}

Now let us consider the partition ${\cal {C}}_3$ of the sphere into 3
zones
of equal volume: $E_1 = \{(\vartheta,\varphi) :\varphi\in [0,2\pi),
\vartheta \in [0,\pi/3]\}$, $E_2 = \{(\vartheta,\varphi) :\varphi\in
[0,2\pi), \vartheta \in (\pi/3,{2\pi}/3]\}$, and $E_3 =
\{(\vartheta,\varphi) :\varphi\in [0,2\pi), \vartheta \in ({2\pi}/3,\pi]\}$.
Set $j = 1/2$. Then applying formula (\ref{ker2}) one can show that the
matrices $D(1), D(2), D(3)$ are given by

\begin{equation}
D(1) = \left(\matrix{1/3 & 2/9 \cr 2/9 & 13/81\cr}\right),~~ D(2) = \left(
\matrix{1/3 & 0 \cr 0 & 1/81\cr}\right),~~ D(3) = \left(\matrix{1/3 & -{2/9}
\cr -{2/9} & 13/81\cr}\right).
\label{IFS7}
\end{equation}
Hence and from (\ref{IFS1}), (\ref{IFS2}) we obtain

\begin{equation}
p_1(\lambda) = 1/3 + (2/9)\lambda, ~~p_2(\lambda) = 1/3, ~~p_3(\lambda) =
1/3 - (2/9)\lambda,
\label{IFS8}
\end{equation}
and

\begin{equation}
F_1(\lambda) = (18 + 13\lambda)/(27 + 18\lambda), ~~F_2(\lambda) = {\lambda}
/27,~~ F_3(\lambda) = (-18 + 13\lambda)/(27 - 18\lambda).
\label{IFS9}
\end{equation}
The set $X = [-1,1]$ fulfills conditions (a)--(d) with the contraction rates
for the maps $F_1, F_2,$ and $F_3$ equal to $c_1 = 1/3$, $c_2 = 1/27$, and
$c_3 = 1/3$, respectively. The support of the attracting invariant measure
$\mu$ presented in Fig.12a is a Cantor-like fractal set.

Now let us consider the case $j=1$ (with the same partition ${\cal {C}}_3$).
Applying formulae (\ref{ker2}), (\ref{IFS1}), and (\ref{IFS2}), we can
compute the maps $p_1, p_2, p_3$, and $F_1, F_2, F_3$, as before. Now, the
set $X = \{(\lambda_1,\lambda_2) : \lambda_1 \in [-1,1],~ {\lambda_1}^2 \le
\lambda_2 \le 1\}$ satisfies conditions (a)--(d). The attractive invariant
set for this IFS is presented in Fig.12b. Also in this case it has a fractal
structure. The view of the middle part of the IFS through a magnifying
glass is shown in Fig.12b to underline its self-similar structure. In the
figure caption we give the values of the CS--measurement entropy obtained with
the aid of the random algorithm.

We have also applied this technique to compute numerically the
CS-measurement entropy for other partitions of the phase space and $j$
ranging from $1/2$ to $10$. For the partition of the sphere into two
hemispheres the results obtained in this way coincide with those received
from the extrapolation of the relative entropies $G_n$ and collected in
Tab.~2.

\section{R\'{e}nyi CS--measurement entropy}

In this section we consider quantities which are natural generalizations of
CS--measurement entropy introduced in sect.~II.C. We assume that ${\cal {C}}$
is a partition of the phase space and the CS--probabilities are given by
(\ref{csprn}). We shall write $H_n$ for $H_n({\cal {C})}$, $G_n$ for
$G_n({\cal {C})}$, and $H_{mes}$ for $H_{mes}({\cal {C})}$. Moreover, we
choose the parameter $\beta > 0$ such that $\beta \neq 1$.

There are at least two different ways of introducing a R\'{e}nyi--type
version of CS--measurement entropy. Firstly, we can define {\sl
CS--measurement entropy of order} $\beta$ as

\begin{equation}
H_{mes}({\beta}) := \limsup_{n\to \infty} {\frac{1 }{n}} H_n({\beta}),
\label{ren1}
\end{equation}
where

\begin{equation}
H_n({\beta}) := {\frac{1 }{1 - \beta }} \ln \Bigl[~\sum_{i_0,
\dots,i_{n-1}=1}^k (P^{CS}_{i_0,\dots,i_{n-1}} )^{\beta} \Bigr] .
\label{ren2}
\end{equation}

On the other hand, using the notion of R\'{e}nyi conditional entropy of
order ${\beta}$ \cite{R61} we can define the quantity

\begin{equation}
G_{mes}(\beta) := \limsup_{n\to \infty} G_n(\beta),
\label{ren3}
\end{equation}
where

\begin{equation}
G_n({\beta}) := \left\{
\begin{array}{lll}
H_1({\beta}) & ~\hbox{for}~ & n = 1; \nonumber \\
{\frac{1 }{1 - \beta}} \ln \Bigl[ {\ \sum_{i_0,\dots,i_{n-1}=1}^k
(P^{CS}_{i_0,\dots,i_{n-1}})^{\beta} (P^{CS}_{i_0,\dots,i_{n-2}})^{1 -
\beta} } \Bigr] & ~\hbox{for}~ & n > 1.
\end{array}
\right.
\label{ren4}
\end{equation}

The quantities $G_n({\beta})$ are the analogues of the relative entropies
considered in sect.~II. Note that
\mbox{$H_n({\beta}) \longrightarrow H_n ~
(\beta \longrightarrow 1)$} and $G_n({\beta}) \longrightarrow G_n ~ (\beta
\longrightarrow 1)$. This justifies the notation $H_n(1) := H_n$, $G_n(1) :=
G_n$, and \mbox{$H_{mes}(1) = G_{mes}(1):= H_{mes}$}.

Contrary to the case $\beta = 1$, the quantities $H_{mes}(\beta)$ and $
G_{mes}(\beta)$ need not be equal in general. On the basis of some numerical
evidences we conjecture that $G_{mes}(\beta) < H_{mes}(\beta)$ for $\beta <
1 $, and $H_{mes}(\beta) < G_{mes}(\beta)$ for $\beta > 1$.

The number $G_n(\beta)$ $(\beta \neq 1)$ can be computed from the
following integral formula analogous with (\ref{IFS5}):

\begin{equation}
G_{mes}(\beta) = {\frac{1 }{1-\beta}} \ln \int_X \sum_{i=1}^{k}
(p_i(x))^{\beta} d{\mu}(x),
\label{ief2}
\end{equation}
where $(X, (F_i)_{i=1}^k, (p_i)_{i=1}^k)$ is the iterated function system
defined in sect.~VI and $\mu$ is the attractive invariant measure for this
system \cite{S96}.

Now let us consider the case of the division of the sphere into two
hemispheres. As in sect.~IV.D we can use the Markovian approximation $G_2({
\beta})$ to evaluate the limiting value $G_{mes}({\beta})$ for large values
of $j$. Similar reasoning leads to the formula

\begin{equation}
G_2({\beta}) = {\frac{1 }{1-\beta}} \ln\Bigl[ {\frac{1+\tau_j^{\beta} }{
(1+\tau_j)^{\beta}}} \Bigr],
\label{ren5}
\end{equation}
where $\beta \neq 1$ and $\tau_j$ is given by formula (\ref{tau}).

The function $G_2$ defined by (\ref{g2j}) and (\ref{ren5}) is continuous.
Moreover, we can compute the limits $G_2({\beta}) \longrightarrow \ln{2}$ ~
\hbox{$(\beta \longrightarrow 0)$} and $G_2({\beta}) \longrightarrow \ln
(1+\tau_j) ~(\beta \longrightarrow \infty)$. Asymptotically (for large $j$)
we obtain

\begin{equation}
G_2(\beta ) \sim \left\{
\begin{array}{lcl}
{\frac{1 }{1 - \beta}} {\frac{1 }{{(2 \pi j)^{{\beta}/2}}}} & ~~~\hbox{for}
~~~ & \beta < 1 \nonumber \\
{\frac{\ln j }{2{(2 \pi j)^{1/2}}}} & ~~~\hbox{for}~~~ & \beta = 1 \\
{\frac{\beta }{\beta - 1}} {\frac{1 }{(2 \pi j)^{1/2} }} & ~~~\hbox{for}~~~
& \beta > 1. \nonumber
\end{array}
\right.
\label{g222}
\end{equation}

In Figs.~13~and~14 we treat the case of the partition of the sphere into two
hemispheres. In Fig.~13 we present the Markovian approximation $G_2$ for
different values of the semiclassical parameter $j$. We see that all the
curves start from the value $\ln{2}$ (topological entropy) and then decrease
when the value of the parameter $\beta$ grows. Moreover, we can observe that
$G_2$ decreases when $j$ increases and converges to $0$ (which is the value
of the classical R\'{e}nyi entropy in this case) if $j$ tends to $\infty$.
In Fig.~14 we compare two versions of the R\'{e}nyi CS--measurement entropy $
H_{mes}({\zeta})$, $G_{mes}({\zeta})$, and the Markovian approximation $G_2({
\zeta})$ for two different values of the parameter $j$. The variable $\zeta
= 4 \arctan(\beta)/\pi$ changes from $0$ to $2$, when $\beta$ varies from $0$
to $\infty$. The quality of the Markovian approximation $G_2$ becomes worse
for large values of $\beta$ and $j$, still, it gives an upper bound for the
R\'{e}nyi CS--entropy $G_{mes}$.

\section{Conclusions}

This work has been devoted to the study of the notion of CS--measurement
entropy. We have collected the basic theoretical material in sects.~II.D and
II.E, analysed numerical algorithms for computing CS-measurement entropy in
sects.~IV.B and VI, examined several examples in sects.~IV and V, and
proposed two generalizations of the notion in sect.~VII. The methods
developed here can be used to investigate of the CS--measurement
entropy for a broad class of partitions of the phase space and values of the
semiclassical parameter $j$. The semiclassical limit (large $j$) has been,
as usual, most difficult to treat. Nevertheless, even in this case, we have
obtained several approximate results in sect.~IV.D. We have
restricted our attention to the spin ($SU(2)$) coherent states defined on
the sphere $S^2$. We believe, however, that our approach can be extended to
other phase spaces and to other families of coherent states.

The fact that the measurement entropy $H_{mes}$ can be calculated as the
limit of the relative entropies $G_n$ has played a crucial role in our
analysis. As we have argued, the approach to the limit is
exponential in this case. The rate of convergence seems to be strictly
connected with the limiting value of the sequence: the larger is the entropy
$H_{mes}$, the faster the convergence. A similar dependence was reported for
the KS--entropy of picewise analytic one--dimensional maps by Sz\'epfalusy
and Gy\"orgyi \cite{SG86}. They estimated the decay of the relative
entropies to be $G_n \sim e^{- 2H(3)n}$, where $H(3)$ is the R\'enyi entropy
of order 3. The convergence we have observed for CS-entropies is much
faster.

In \cite{SZ94} we formulated a general programme
 for analysing {\sl quantum
chaos} in terms of CS--entropy. Here, we have studied CS--measurement
entropy only, that is, the CS--entropy of the identity operator, which
measures the randomness coming from the process of approximate sequential
quantum measurement. Still, our main purpose is to study {\sl CS--dynamical
entropy}, which is connected only with the unitary dynamics of the quantum
system and is defined as the difference of two quantities: the CS--entropy of
the given unitary operator and the CS--measurement entropy (see formula
(\ref{dyn2})). The precise analysis of the notion of CS--measurement entropy
is the first essential stage in performing this task. We expect that the
methods elaborated here can also be used in the investigation of the
CS--entropy for an arbitrary unitary map $U$, and so, in studying
CS--dynamical entropy. The main difficulty in extending our approach to the
general case is that we have to deal with much larger matrices,
notwithstanding, the numerical algorithms can be managed in much the same
way. In a forthcoming publication we shall try to calculate the CS--dynamical
entropy for quantized regular and chaotic maps.

In this work we have presented an effective method of computing the
dynamical entropy of a system via {\sl iterated function systems}. Although
this technique has been applied here only in calculations of the
CS-measurement entropy, we believe that it may be useful for computing the
CS--dynamical entropy of quantum systems, as well as the Kolmogorov--Sinai
entropy of classical systems.

\section*{Acknowledgements}

We thank Mark Fannes, Piotr Garbaczewski, Denes Petz, Franklin Schroeck, and
Tomasz Zastawniak for enlightening discussions. This work was supported by
the Polish KBN grant 2~P03A~060~09.

\appendix

\section{Bounds for CS--measurement entropy}

\label{s:appendix A}

{\sl Proof of the inequalities (\ref{bound1}).}

\vskip 0.2cm
We assume that ${\cal C}$ denotes a finite partition of the
phase space and we put $H_{mes} := H_{mes}({\cal {C}})$,
$H_n := H_n({\cal {C}})$, and $G_n := G_n({\cal {C}})$.

It follows from the general theory of dynamical entropy \cite{ME81,K88} that
the sequence $\{H_n\}_{n \in N}$ is subadditive, i.e.,

\begin{equation}
H_{n+l} \le H_n + H_l ~~~{\rm for}~~ n,l \in {\N2}.
\label{subadd}
\end{equation}
Let now $n,l \in {\N2}$, $i_0, \dots, i_{n+l-1} = 1, \dots, k$. Then from
(\ref{csprn}) we deduce

\begin{eqnarray}
\lefteqn{ P_{i_0,\dots,i_{n+l-1}}^{CS} = \int_{E_{i_0}} dx_0 \cdots
\int_{E_{i_{n-1}}} dx_{n-1} {\displaystyle{\prod_{u=1}^{n-1}}}
{K(x_{u-1},x_u)} \int_{E_{i_n}} dx_n~{K(x_{n-1},x_n)}~ \times} \hspace{
2.0cm}  \nonumber \\
& & \times \int_{E_{i_{n+1}}} dx_{n+1} \cdots \int_{E_{i_{n+l-1}}}
dx_{n+l-1} {\displaystyle{\prod_{u=n+1}^{n+l-1}}} K(x_{u-1},x_u) ~\le ~
P_{i_0,\dots,i_{n-1}}^{CS} \cdot K_0 \cdot P_{i_n,\dots,i_{n+l-1}}^{CS}.
\label{app1}
\end{eqnarray}
Taking the logarithms of both sides of (\ref{app1}), multiplying them by $-
P_{i_0,\dots,i_{n+l-1}}^{CS}$, and summing over
\hbox{$i_0, \dots, i_{n+l-1}
= 1, \dots, k$} we get

\begin{equation}
H_{n+l} \ge H_n - \ln{K_0} + H_l ~~~{\rm for}~~ n,l \in {\N2}.
\label{subaddc}
\end{equation}
Combining (\ref{subadd}) with (\ref{subaddc}) and dividing the expressions
by $n$ we have

\begin{equation}
{\frac{1 }{n}} H_n - {\frac{1 }{n}} \ln{K_0} \le {\frac{1 }{n}} {(H_{l+n} -
H_{l})} \le {\frac{1 }{n}} H_n,
\label{app2}
\end{equation}
and so

\begin{equation}
{\frac{1 }{n}} H_n - {\frac{1 }{n}} \ln{K_0} \le {\frac{1 }{n}} {
\sum_{i=1}^n {G_{l+i}}} \le {\frac{1 }{n}} H_n.
\label{app3}
\end{equation}
Letting $l \to \infty$ we obtain the desired conclusion.

\section{Convergence rate of partial entropies}

\label{s:appendix B}

{\sl Proof of property II.D.(4).}

\vskip 0.2cm
We follow the notation of Appendix A. From II.D.(1) we get

\begin{equation}
{\frac{1 }{n}}{H_n} = {\frac{1 }{n}}{\sum_{i=1}^n {G_i}} \ge {\frac{1 }{n}}{
H_1} + {\frac{{n-1}}{{n}}}{H_{mes}}.
\label{app4}
\end{equation}
Hence

\begin{equation}
{\frac{1 }{n}}{H_n} - H_{mes} \ge {\frac{{H_1 - H_{mes}} }{{n}}} > 0.
\label{app5}
\end{equation}
On the other hand (\ref{bound1}) implies

\begin{equation}
{\frac{1 }{n}}{H_n} - H_{mes} \le {\frac{{\ln{K_0}} }{{n}}}.
\label{app6}
\end{equation}
Combining (\ref{app4}) and (\ref{app5}) we get the required result.

\section{Formula for the second order CS-probabilities}

\label{s:appendix C}

{\sl Proof of formula (4.6).}

\vskip 0.2cm
Set $2j = M$. Then, from (\ref{prob4}) and (\ref{ker2}) we have

\begin{eqnarray}
P^{CS}_{+-} & = \int_{0}^{1}\! {\frac{dt }{{2}}} \int_{-1}^{0}\! {\frac{ds }{
{2}}} {\tilde{K}}(t,s) = {\frac{M+1 }{4^{M+1}}} {\displaystyle{\
\sum_{q=0}^{M}}} \Bigl( {M \atop q} \Bigr)^2 \int_{0}^{1} dt
{(1+t)}^q {(1-t)}^{M-q} \int_{-1}^{0} ds {(1+s)}
^q {(1-s)}^{M-q}  \nonumber \\
& = {\frac{1}{{(M+1) 4^{M+1}}}} {\displaystyle{\ \sum_{q=0}^{M}}} R_{q}^{M}
R_{M-q}^{M},
\label{app7}
\end{eqnarray}
where
\begin{equation}
R_{p}^{M} := (M+1) \Bigl({M \atop p} \Bigr) \int_{0}^{1} dt
{(1+t)}^p {(1-t)}^{M-p}.
\label{app8}
\end{equation}
Now we need the following two lemmas, which we shall prove later.

\vspace{0.2cm}
{\bf Lemma 1.}
\begin{equation}
R_{p}^{M} = \sum_{s=0}^{p} \Bigl({M+1 \atop s} \Bigr);
\label{app9}
\end{equation}
and

\vspace{0.2cm}
{\bf Lemma 2.}
\begin{equation}
\sum_{q=0}^M R_q^M R_{M-q}^M = (2M+1) \Bigl({2M  \atop M} \Bigr).
\label{app10}
\end{equation}
Combining (\ref{app7}) and (\ref{app10}) we get

\begin{equation}
P^{CS}_{+-} = {\frac{1}{{(M+1) 4^{M+1}}}} (2M+1) \Bigl( {2M \atop M}
\Bigr) = \Bigl( {2M+1 \atop M} \Bigr) {\frac{1}{{4^{M+1}}}},
\label{app11}
\end{equation}
which establishes the formula.

\vspace{0.2cm} \noindent
{\sl Proof of Lemma 1.} We proceed by induction. Clearly, $R_{0}^{M} = 1$.
Assuming (\ref{app9}) to hold for $p$, we shall prove it for $p+1$. We have

\begin{equation}
R_{p+1}^{M} = (M+1) \Bigl( {M \atop p+1} \Bigr) \int_{0}^{1} dt
{(1+t)}^{p+1} {(1-t)}^{M-p-1}.
\label{app12}
\end{equation}
Integrating by parts we obtain

\begin{equation}
R_{p+1}^{M} = {\frac{M+1 }{M-p}} \Bigl(
{M \atop p+1}\Bigr) + {\frac{{(M+1)(p+1)}}{M-p}} \Bigl(
{M \atop p+1} \Bigr) \int_0^1 dt (1+t)^p (1-t)^{M-p}.
\label{app13}
\end{equation}
By the induction assumption

\begin{equation}
R_{p+1}^{M} = \Bigl({M+1 \atop p+1} \Bigr) + R_p^M,
\label{app14}
\end{equation}
which completes the proof.

\vspace{0.2cm} \noindent
{\sl Proof of Lemma 2.} Applying Lemma~1 we deduce that

\begin{eqnarray}
\sum_{q=0}^M R_q^M R_{M-q}^M & = & \sum_{\stackrel{\scriptstyle{s,l=0;}}{s+l
\le M}}^M \Bigl( {M+1 \atop s} \Bigr) \Bigl({M+1 \atop l} \Bigr)
((M+1)-(s+l))  \nonumber \\
& = & \sum_{r=0}^{M+1} \sum_{s=0}^{r} \Bigl( {M+1 \atop s} \Bigr)
\Bigl( {M+1  \atop r-s} \Bigr) ((M+1)-r).
\label{app15}
\end{eqnarray}
Using the well--known combinatorial identities

\begin{equation}
\sum_{s=0}^{L} \Bigl({L \atop s} \Bigr) \Bigl( {L \atop r-s} \Bigr) =
\Bigl( {2L \atop r} \Bigr);
\label{app16}
\end{equation}

\begin{equation}
\sum_{r=0}^{L} \Bigl( {2L \atop r} \Bigr) = {\frac{1 }{2}} \Bigl( 4^L
+ \Bigl( {2L \atop L} \Bigr) \Bigr);
\label{app17}
\end{equation}

\begin{equation}
\sum_{r=0}^L \Bigl( {2L \atop r} \Bigr) r = {\frac{L }{2}} {4^L},
\label{app18}
\end{equation}
we conclude that

\begin{eqnarray}
\sum_{q=0}^{M} R_{q}^{M} R_{M-q}^{M} & = & \sum_{r=0}^{M+1} \Bigl(
{2M+2 \atop r} \Bigr) ((M+1) - r) =  \nonumber \\
& = & {\frac{{M+1} }{2}} \Bigl( {2M+2 \atop M+1} \Bigr) = (2M+1)
\Bigl( {2M \atop M}\Bigr),
\label{app19}
\end{eqnarray}
which proves the lemma.

\newpage

\begin{figure}[tbp]
\caption{Scheme of the first three periods of the time evolution of the
dynamical system. The unitary quantum map $U$ describes the evolution of the
system during each period, after which an act of an approximate measurement
takes place.}
\label{ff1}
\end{figure}

\begin{figure}[tbp]
\caption{Spherical representation of the spin coherent state $|\vartheta,
\varphi \rangle$ generated by the unitary rotation operator $R(\vartheta,
\varphi)$.}
\label{ff2}
\end{figure}

\begin{figure}[tbp]
\caption{CS--measurement entropy $H_{mes}$ for two hemispheres as a function
of the quantum number $j$. Circles represent numerical results, while the
solid line stands for an upper bound $G_2$ given by
 (\protect\ref{g2j}).}
\label{ff3}
\end{figure}

\begin{figure}[tbp]
\caption{Partition of the sphere divided along a parallel $\Theta_c$ into
two connected cells.}
\label{ff4}
\end{figure}

\begin{figure}[tbp]
\caption{Partial entropy $H_n/n$ for the partition presented in Fig.4 as
a function of the parameter $\cos \Theta_c$ ($n=8$). The values of $j$ are
given in the picture.}
\label{ff5}
\end{figure}

\begin{figure}[tbp]
\caption{Partial entropies $H_n/n$ and measurement entropy $
H_{mes}=\lim_{n\to\infty} H_n/n$ for the partition presented in Fig.4 with
a) $j=1/2$, and b) $j=5$.}
\label{ff6}
\end{figure}

\begin{figure}[tbp]
\caption{Partition of the sphere into a) two cells plotted out by parallels $
\Theta_d$ and $\pi-\Theta_d$: a spherical zone and the union of two
spherical segments; b) two disconnected cells created by the equator and the
spherical wedge of the radian measure $\Phi_d$.}
\label{ff7}
\end{figure}

\begin{figure}[tbp]
\caption{Dependence of CS-measurement entropy on $\cos \Theta_d$ for the
partition of the sphere into two cells presented in Fig.~7a and $j=1/2$.}
\label{ff8}
\end{figure}

\begin{figure}[tbp]
\caption{Dependence of CS-measurement entropy on $\cos \Phi_d$ for the
partition of the sphere into two cells presented in Fig.~7b and $j=1/2$.}
\label{ff9}
\end{figure}

\begin{figure}[tbp]
\caption{Partition of the sphere into $k$ zones of equal volume.\hfill}
\label{ff10}
\end{figure}

\begin{figure}[tbp]
\caption{CS--measurement entropy $H_{mes}$ as a function of the number of
cells $k$ for $j=1/2$ (circles). Solid line
represents the function $\ln{k}$.
The inset shows the dependence of
 $H_{max}:= \ln{k} - H_{mes}$ on $k^{-2}$
obtained for $k=30,\dots,1000$.}
\label{ff11}
\end{figure}

\begin{figure}[tbp]
\caption{The attracting invariant set for the IFS generated by the partition
of the sphere into three zones of equal volume for a) $j=1/2$, b) $j=1$. The
values of the CS--measurement entropy computed by the random iterated
algorithm are a)~$H_{mes} = 1.05306
\dots$, b)~$H_{mes} = 0.99220
\dots$.}
\label{ff12}
\end{figure}

\begin{figure}[tbp]
\caption{Markovian approximation $G_2$ of the R\'{e}nyi CS-measurement
entropy as a function of the parameter $\beta$ for the partition of the
sphere into two hemispheres and selected values of the quantum number $j$
labeling the curves.}
\label{ff13}
\end{figure}

\begin{figure}[tbp]
\caption{Two versions of the R\'{e}nyi CS-measurement entropy $H_{mes}$, $
G_{mes}$, and the Markovian approximation $G_2$ as a function of the
rescaled parameter $\zeta = 4\arctan(\beta)/\pi$ for $j=1$ and $j=5$, in the
case of the partition of the sphere into two hemispheres.}
\label{ff14}
\end{figure}

\newpage
\begin{table}[tbp]
\caption{Coherent states probabilities $P^{CS}$ for the division of the
sphere into two hemispheres: the number of measurements $n=2,3,4$, the
quantum number $j=1/2,1,3/2$, with the limit $j\to\infty$.}
\label{tab1}\vskip 1.0cm
\begin{tabular}{|c|c|c|c|c|c|c|c|}
& $j=$ & 0 & 1/2 & 1 & 3/2 & $\dots$ & $j\to \infty$ \\ \hline
n=2 & $P^{CS}_{++}=P^{CS}_{--}$ & 1/4 & 5/16 & 11/32 & 93/256 & \dots & 1/2
\\
n=2 & $P^{CS}_{+-}=P^{CS}_{-+}$ & 1/4 & 3/16 & 5/32 & 35/256 & \dots & 0 \\
\hline
n=3 & $P^{CS}_{+++}=P^{CS}_{---}$ & 1/8 & 19/96 & 31/128 & 975/3584 & \dots
& 1/2 \\
n=3 & $P^{CS}_{++-}=P^{CS}_{--+}$ & 1/8 & 11/96 & 13/128 & 327/3584 & \dots
& 0 \\
n=3 & $P^{CS}_{-++}=P^{CS}_{+--}$ & 1/8 & 11/96 & 13/128 & 327/3584 & \dots
& 0 \\
n=3 & $P^{CS}_{+-+}=P^{CS}_{-+-}$ & 1/8 & ~7/96 & ~7/128 & 163/3584 & \dots
& 0 \\ \hline
n=4 & \ $P^{CS}_{++++}=P^{CS}_{----}$\  & ~1/16~ & ~289/2304~ & ~5609/32768~
& ~8225957/40140800~ & $\dots$ & 1/2 \\
n=4 & $P^{CS}_{+++-}=P^{CS}_{---+}$ & ~1/16~ & ~167/2304~ & ~2327/32768~ &
~2694043/40140800~ & $\dots$ & 0 \\
n=4 & $P^{CS}_{+---}=P^{CS}_{-+++}$ & ~1/16~ & ~167/2304~ & ~2327/32768~ &
~2694043/40140800~ & $\dots$ & 0 \\
n=4 & $P^{CS}_{++--}=P^{CS}_{--++}$ & ~1/16~ & ~161/2304~ & ~2153/32768~ &
~2429093/40140800~ & $\dots$ & 0 \\
n=4 & $P^{CS}_{++-+}=P^{CS}_{--+-}$ & ~1/16~ & ~103/2304~ & ~1175/32768~ &
~1233307/40140800~ & $\dots$ & 0 \\
n=4 & $P^{CS}_{+-++}=P^{CS}_{-+--}$ & ~1/16~ & ~103/2304~ & ~1175/32768~ &
~1233307/40140800~ & $\dots$ & 0 \\
n=4 & $P^{CS}_{+--+}=P^{CS}_{-++-}$ & ~1/16~ & ~~97/2304~ & ~1001/32768~ &
~~968357/40140800~ & $\dots$ & 0 \\
n=4 & $P^{CS}_{+-+-}=P^{CS}_{-+-+}$ & ~1/16~ & ~~65/2304~ & ~~617/32768~ &
~~592293/40140800~ & $\dots$ & 0
\end{tabular}
\end{table}

\vskip 1.0cm
\begin{table}[tbp]
\caption{Partial entropy ${H_n}/n$ and relative entropy $G_n$ for the
partition ${\cal {C}} = \{E_+,E_-\}$, the quantum number $j=1/2$ and $j=10$,
and the number of measurements $n=1,\dots,15$ with an extrapolation to $n
\to \infty$.}
\label{tab2}\vskip 1.0cm 
\begin{tabular}{|c||c|c||c|c||}
& ${H_n}/n$ & $G_n$ & ${H_n}/n$ & $G_n$ \\ \hline
~n~ & \multicolumn{2}{c||}{$j=1/2 $} & \multicolumn{2}{c|}{$j=10$} \\ \hline
\ 1 \  & 0.693147180559 & 0.693147180559 & \ 0.6931471\  & \ 0.6931471 \  \\
\ 2 \  & \ 0.677355209358\  & \ 0.661563238157\  & \ 0.5323993\  & \
0.3716514\  \\
\ 3 \  & \ 0.672009066259\  & \ 0.661316780060\  & \ 0.4734456\  & \
0.3555383\  \\
\ 4 \  & \ 0.669335388698\  & \ 0.661314356017\  & \ 0.4429253\  & \
0.3513642\  \\
\ 5 \  & \ 0.667731177545\  & \ 0.661314332934\  & \ 0.4243127\  & \
0.3498623\  \\
\ 6 \  & \ 0.666661703407\  & \ 0.661314332713\  & \ 0.4118012\  & \
0.3492438\  \\
\ 7 \  & \ 0.665897793307\  & \ 0.661314332711\  & \ 0.4028258\  & \
0.3489737\  \\
\ 8 \  & \ 0.665324860733\  & \ 0.661314332711\  & \ 0.3960793\  & \
0.3488532\  \\
\ 9 \  & \ 0.664879246508\  & \ 0.661314332711\  & \ 0.3908259\  & \
0.3487993\  \\
\ 10\  & \ 0.664522755128\  & \ 0.661314332711\  & \ 0.3866209\  & \
0.3487751\  \\
\ 11\  & \ 0.664231080363\  & \ 0.661314332711\  & \ 0.3831794\  & \
0.3487645\  \\
\ 12\  & \ 0.663988018059\  & \ 0.661314332711\  & \ 0.3803111\  & \
0.3487597\  \\
\ 13\  & \ 0.663782349955\  & \ 0.661314332711\  & \ 0.3778839\  & \
0.3487577\  \\
\ 14\  & \ 0.663606063009\  & \ 0.661314332711\  & \ 0.3758034\  & \
0.3487568\  \\
\ 15\  & \ 0.663453280989\  & \ 0.661314332711\  & \ 0.3740002\  & \
0.3487564\  \\
\ $\vdots$ \  & $\ \vdots$ & \ $\ \vdots$ \  & \ $\vdots$ \  & \ $\vdots$ \
\\ \hline
\ $\infty$\  & \ 0.6613 \  & \ 0.661314332711 \  & \ 0.3488 \  & \
0.3487560 \
\end{tabular}
\end{table}

\end{document}